\documentstyle[amssymb,12pt]{article}

\input{tcilatex}

\begin{document}

\baselineskip=14pt

\begin{quote}
{\bf \ }
\end{quote}

\begin{center}
{\Huge Symbology from set theory applied to ecological systems: Gause's
exclusion principle and applications}
\end{center}

\[
\]

\begin{center}
{\Large \ J. C. Flores}$^{a,b}$

$^{a}$Universidad de Tarapac\'{a}, Departamento de F\'{i}sica, Casilla 7-D,
Arica-Chile

$^{b}$Centro de Estudios del Hombre en el Desierto, CIHDE, Casilla 7-D,
Arica-Chile
\end{center}

\[
\]

{\bf Abstract: } We introduce a symbolic representation like set theory to
consider ecologic interactions between species (ECOSET). The ecologic
exclusion principle (Gause) is put in a symbolic way and used as operational
tool to consider more complex cases like interaction with sterile species
(SIT technique), two species with two superposed sources (niche
differentiation) and N+P species competing by N resources, etc. Displacement
(regional or characters) is also considered by using this basic tool. Our
symbolic notation gives us an operative and easy way to consider elementary
process in ecology. Some experimental data (laboratory or field) for
ecologic process are re-considered under the optic of this set-theory.

\[
\]


\[
\]

\bigskip\ 
\[
\]

{\Large I Introduction.-}

\[
\]

Interactions between species in ecology is after some time the object of
study of mathematical branches (Gertsev et al (2004) and references
therein). For instance, mathematical models for predator-prey are usually
treated with nonlinear coupled differential equations like Lotka-Volterra
(Begon et al (1999), Murray (1993), and references therein). It is true
that, as a general rule, modellation of ecological systems is a difficult
task since complexity in biological sciences is almost always present. For
instance, the more known application of Lotka-Volterra system, that is the
Hare-Lynx predator-prey data recorder by Hudson Bay Company in 1953,
presents some troubles. The expected periodic solution between prey and
predator is not determined from data (Murray (1993)). The reasons are not
clear but one can expect a complex dynamics for real predator-prey system in
a not isolated region. Very refined experiment in laboratories are actually
realized (Costantino et al (1995)), nevertheless the more impressive
laboratory experiments were carried-out some time ago by ecologist G. F.
Gause with protozoan {\it Paramesium }species. These experiments have
determined some general principles applied to inter-competition and
coexistence in ecology. The more basic for us is the statement that: when we
have two species competing by (exactly) the same niche then one of them
disappear (Gause (1934)). That is, finally, one species full the niche. The
statement is very restrictive since it requires quite special conditions
(section IV) and in this paper it will be used as a basic principle to
consider more complex situations.

Mathematical models like Lotka-Volterra for two species in competition
predict coexistence for some range of parameters (high intraspecific
competition). In fact, any mathematical system with atractors (out-side of
the axis) in this two dimensional phase-space makes the same prediction. So,
for us this mathematical result is not proof of real violation of the
principle. Gause's exclusion principle will be assumed for interspecific
competition as a basic statement in this paper. Nevertheless, we accord that
the distinction between the minimum amount of niche differentiation (in real
ecological system) to produce its break is a difficult point to consider in
practice. For an appropriate discussion see reference (Begon et al (1999)).

In this paper we will consider a mathematical modeling for ecological
systems but using a symbology, called ECOSET, similar to basic set theory.
This schema has the advantage of a condensed notation for a variety of
ecological interacting systems. There are some similarities with usual set
theory but also some differences. For instance, Gause's exclusion principle
(16) has not equivalence in the usual set theory. In our construction,
Gause's exclusion principle will be used as a basic operational tool to
consider complex situations like more than one species and more than one
resource. A great part of field and laboratory examples found in this
article were hold from reference Begon et al (1999).

In this paper, species will be represented by capital letter like $A$, $B,$%
...etc. We will use the symbol $S$ only for a primary source at the basis
(consume) of a ecological chain. We will give to $S$ a stronger sense: it
will design a defined ensemble of sources which make viable the development
of species. Namely, it is considered in the sense of an ecological niche. We
assume that this primary resource is auto-sustained or externally sustained.
The absence of species in a given region will be denoted by $\phi $ in
analogy with the symbol of usual set theory.

\[
\]

{\Large II Symbology for depredation (\TEXTsymbol{>}) and basic
definitions.- }

\[
\]

Consider a primary resource $S$ and species $A,B,C...\phi $. The notation

\begin{equation}
A>B\text{ \ (}B\text{ \ consumes }A\text{),}
\end{equation}
means: species $B$ exploits species $A$ as a resource in a sense of
depredation. In this way, a basic chain becomes for instance 
\begin{equation}
S>A>B>\phi \ \ (\text{a basic depredation chain }),
\end{equation}
namely, species $B$ exploits $A$ as a resource and, $A$ exploits the primary
resource $S$. Note that the species $\phi $ at the end of the chain means
that $B$ is not prey for others.

For two no-interacting species $A$ and $B$ (also for ecological process) in
a given ecological region we write $A\oplus B$. When two species ($A $ and $%
B $) consume the same primary resource ($S$), in a not depending way, we
write

\begin{equation}
S>\left( A\oplus B\right) ,\text{ \ (independent depredation).}
\end{equation}

When two species ($A$ and $B$) exploit the same resource ($S$) in a
interdependent way we write

\begin{equation}
S>\left( A\otimes B\right) \text{ \ (interdepending depredation).}
\end{equation}
The notation $S>A\otimes B$ (without braces) means: species $A$ consumes $S$
with the help of species $B$.

Note that when one of the interdependent species if $\phi $ then, after a
time, no depredation on $S$ must occur. In a symbolic way,

\begin{equation}
\left\{ S>\left( A\otimes \phi \right) \right\} \Rightarrow S>\phi ,
\end{equation}
where the symbol $\Rightarrow $ has a temporal interpretation or, it defines
a temporal direction. Namely, if we have the ecological system $X$ then,
after a time we have $Y$ ($X$\ $\Rightarrow Y$). For instance, the
ecological proposition: {\it when one species }$A${\it \ has not source for
depredation then it dies}, could be written as:

\begin{equation}
\left\{ S>\phi >A\right\} \Rightarrow \left\{ S>\phi \right\} .
\end{equation}
\ 

We also define the symbol $\Leftrightarrow $ which will be interpreted as
equivalence between ecological process. For instance, in the process (3) we
always \ assume implicitly the equivalence: 
\begin{equation}
\left\{ S>\left( A\oplus B\right) \right\} \Leftrightarrow \left\{ \left(
S>A\right) \oplus \left( S>B\right) \right\} .
\end{equation}

Two notes: (a) A ecological process like $S>A>B$ does not mean that $S>B$.
If it is true that also $B$ consumes $S$ we must write $\left( S>A>B\right)
\oplus \left( S>B\right) $. (b) We have the equivalent notation $%
A>B\Leftrightarrow B<A$.

As a field example consider the food web with four trophic levels from New
Zealand stream community (Begon et al (1999) page 836). This ecological
systems is composed by Algae ($A$), Herbivorous insects ($H$), Predatory
insects ($P$) and Brown trout ($B$). The web food is :

\begin{equation}
\left\{ \ A>H>P>B\right\} \oplus \left\{ A>H>B\right\} ,
\end{equation}
and then

\begin{equation}
\ (8)\Leftrightarrow A>\left\{ \left( H>P\right) \oplus H\right\} >B.
\end{equation}
\ So, $A$ \ is a primary resource and $B$ the final predator. Note that if $%
P\Rightarrow \phi $ (extinction), the web food does not disappear completely
since $\left\{ A>H>B\right\} $. As expected, biodiversity leads stability of
ecological systems.

As another field example consider the ecological trophic level at the Lauca
National Park (Arica-Chile). There is a biodiversity group represented by
ancient flora and fauna under extreme climatic condition (3.0 to 4.5 Km of
altitude). The highly adapted species conform almost a close system.
Particularly we have a partial food-chain composed by different species
like: basic herbs, including the so-called {\it bofedal}, ($H$); a kind of
camel called Vicunas ($VC$); a kind of rodent called Vizcacha ($VZ$); two
predators (Puma ($P$) and Zorro ($Z$)). Also we have the carrion-eat species
condor ($K$). A basic process of Lauca National Park is represented by 
\begin{equation}
\ H>\left( VC\oplus VZ\right) >\left\{ \left( P\oplus Z\right) \oplus
K\otimes (P\oplus Z)\right\} .
\end{equation}
In fact, species $P$ and $Z$ are natural competitors in this region.
Naturally, the web is more complex of that represented by (10), for instance
species $K$ also depends on natural dead of $VC$ and $VZ$; but is only a
basic notation example for us.

\[
\]

{\Large III\ Symbology for competition (}$\supset \subset ${\Large ) and
basic definitions.- } 
\[
\]

To consider species in competition (no depredation) we will use the symbol $%
\supset \subset $ (see later). Here we give some basic definitions, for
instance, consider species $A$ and $B$ in struggle for some source like
water, space, etc. The symbol: 
\begin{equation}
A\supset B\text{ \ (}A\text{ is perturbed by }B\text{),}
\end{equation}
means that species $B$ perturbs (interferes) $A$. Note that for depredation
we use other symbol ($>$). The above process could also written as $B\subset
A$ or in the equivalence language

\begin{equation}
A\supset B\text{ \ }\Leftrightarrow B\subset A\text{.}
\end{equation}
The symbols $\oplus ,\otimes ,\Rightarrow $\ and $\ \Leftrightarrow $ are
used in the same way that in the above section.

With this basic definitions we can represent competition between species. In
fact, if $A$ and $B$ are two species in competition (no depredation) we write

\begin{equation}
\left( A\supset B\right) \oplus \left( B\supset A\right) ,\text{ \ (}A\text{
and }B\text{ compete).}
\end{equation}

By simplicity we will use the alternative symbol $\supset \subset $ for
competition. Namely, 
\begin{equation}
\left\{ \left( A\supset B\right) \oplus \left( B\supset A\right) \right\}
\Leftrightarrow \left\{ A\supset \subset B\right\} ,\text{ \ (symbol for
competition).}
\end{equation}

Before to ending this section we give a useful definition which will be used
\ in some cases. The notion of ``potential competitors'' is related to two
species who put together then compete. In our symbolic notation, it could be
written as

\begin{equation}
\left\{ S>\left( A\oplus B\right) \right\} \Rightarrow \left\{ S>\left(
A\supset \subset B\right) \right\} \text{ (potential competitors).}
\end{equation}

\[
\]

{\Large IV Gause's exclusion principle for interspecific competition and
symbolic notation.- }

\[
\]

Gause's exclusion principle (or competitive exclusion principle) in ecology
states that: {\it when we have two species }$A${\it \ and }$B${\it \ which
compete (interspecific competition) for the same invariable ecological
primary resource }$S${\it \ (realized niche), then one of them disappear}
(Begon et al (1999), Gause (1934), Hasting (1996), Flores (1998)). It is
important to note that Gause's exclusion principle holds when no migration,
no mutation and no resource differentiation exist in the ecological systems.
Note that it refers to interspecific competition. The case of intraspecific
competition will be touched briefly in section IX. The principle assures
that the more stronger species in the exploitation of primary resource
survives. Applications could be found in many text of ecology. A direct
application of this principle to Neanderthal extinction in Europe could be
found in reference (Flores (1998)).

In our symbolic notation the principle could be written as:

\begin{equation}
\left\{ S>\left( A\supset \subset B\right) \right\} \text{ }\Rightarrow
\left\{ \left( S>A\right) \text{ \ or }\left( S>B\right) \right\} ,\text{ \
(Gause).}
\end{equation}

The above statement (16) will be a basic operational tool to consider more
general cases or application like two sources and two predators, or more
general. So, (16) is our start-point. The logic operator $or$ (some times
written as $\vee $) is the usual exclusion symbol in set theory.

The more famous example of exclusion comes from the classic laboratory work
of ecologist G. F. Gause (1934), who considers two type of {\it Paramecium},
namely, {\it P. caudatum} and {\it P. aurelia}. Both species grow well alone
and reaching stable carrying capacities in tubes of liquid medium and
consuming bacteria. When both species grow together, {\it P. caudatum}
declines to the point of extinction and leaving {\it P. aurelia} in the
niche.

As said before, other examples could be found in literature. For instance,
competition between {\it Tribolium confusum }and {\it Tribolium castaneum}
where one species is always eliminated when put together (Park (1954)).

\[
\]

{\Large V Application: interaction with sterile individuals and eradication
(SIT).-}

\[
\]

A corollary of the above principle can be found when one of the species in
competition is sterile. In fact, we define a sterile specie $M$ as a species
which exploits a resource $S$ and then disappear. Namely,

\begin{equation}
\left\{ S>M\right\} \Rightarrow \left\{ S>\phi \right\} ,\text{ \ (sterile
species).}
\end{equation}

Now, \ we consider this definition together to the exclusion principle. Let $%
A$ \ be a species which exploits the resource $S$, and let $M$ \ be a
sterile species \ introduced which exploits the same resource. \ The
application direct of the principle (16), and definition (17), tell us that 
\begin{equation}
S>\left( A\supset \subset M\right) \Rightarrow \left( S>A\right) \text{ \ or
\ }\left( S>M\right) ,\text{ \ (Gause applied).}
\end{equation}
and then, 
\begin{equation}
\Rightarrow \left\{ S>A\right\} \text{ \ or \ }\left\{ S>\phi \right\} ,%
\text{ \ (}M\text{ is sterile ). }
\end{equation}
Putting together (18) and (19), we have the ecological process:

\begin{equation}
\left\{ S>\left( A\supset \subset M\right) \right\} \Rightarrow \left\{
\left( S>A\right) \text{ \ or }\left( S>\phi \right) \right\} ,\text{ \
(Gause for sterile),}
\end{equation}
so, at least one species of both disappear \ and then there is the
possibility of total extinction in the niche ($S>\phi $). The known SIT
(Sterile Insect \ Technique, Barclay (2001)) uses this principle to
eradicate undesirable insects. In fact, sterile insects compete with native
ones for a source and there is \ the possibility of total extinction
(eradication, $S>\phi $). \ The fruit flies ({\it medfly}) eradication
program carried out in many regions of the world, for instance in
Arica-Chile, \ could be understand partially with the above results \
(Flores (2000) and (2003)). \ If \ $S$ \ is the female-native \ group then
the native male group $A$ and the sterile \ male group $\ M$ compete by the
\ ``resource $S$''. In this way using (20) there is the possibility of $%
S>\phi $ corresponding in this case to extinction of all type of male and
then the wild species disappear.

\[
\]

{\Large VI \ Application: \ two species, two resources, and niche
differentiation.-}

\[
\]

As other application of our symbology for the principle of Gause, consider
two resources $S_{1}$ \ and $S_{2}$ and two species $A$ and $B$ in
competition by these resources. Namely, consider the ecological \ systems
where

\begin{equation}
\left( S_{1}\oplus S_{2}\right) >\left( A\supset \subset B\right) ,
\end{equation}
or

\begin{equation}
(21)\Leftrightarrow \left\{ S_{1}>\left( A\supset \subset B\right) \right\}
\oplus \left\{ S_{2}>\left( A\supset \subset B\right) \right\} ,
\end{equation}
the equivalence becomes since \ both species \ consume any of two resources.
From \ the exclusion principle (16), we have

\begin{equation}
(22)\Rightarrow \left\{ S_{1}>A\text{ \ or \ }S_{1}>B\right\} \oplus \left\{
S_{2}>A\text{ \ or \ }S_{2}>B\right\} ,\text{ (Gause applied),}
\end{equation}
and the four final possibilities:

\[
\]

(a) $\left\{ S_{1}\oplus S_{2}\right\} >A.$ Species \ $A$ exterminates $B$.

(b) $\left\{ S_{1}\oplus S_{2}\right\} >B.$ Species \ $B$ exterminates $A$.

(c) $\left\{ S_{1}>A\right\} \oplus \left\{ S_{2}>B\right\} $. Species $\ A$
exploits $S_{1}$ \ and $B$ exploits $S_{2}$.

(d)$\left\{ S_{1}>B\right\} \oplus \left\{ S_{2}>A\right\} $. Species $A$
exploits $S_{2}$ \ and $B$ exploits $S_{1}$. \ \ \ \ \ \ \ \ \ \ \ \ \ \ \ \
\ \ \ \ \ \ \ \ \ \ \ \ \ \ \ \ \ \ \ \ \ \ \ \ \ \ \ \ \ \ \ \ \ \ \ \ \ \
\ \ \ \ \ \ \ \ \ \ \ \ \ \ \ \ \ \ \ \ 

\[
\]
The last two possibilities (c) and (d) \ tell us that both species could
survive \ by resources exploitations in a differential (partitioned) way.
Some time this coexistence is considered a violation of Gause's exclusion
principle; but it is not. In fact, we have \ more than one resource
(realized niche).

The behavior \ found in the above process (a-d), has been observed in
laboratories (see Begon et al (1999) page 311, or Tilman (1977)) where two
diatom species ({\it Asterionella formosa }and{\it \ Cyclotella meneghimiana}%
) compete by silicate ($S_{1}$) and phosphate ($S_{2}$) as elementary
resources. In fact, for different proportions of this components one can see
extermination or stable coexistence (Tilman (1977)). This is a valuable
laboratory experimental example which support our \ theory as a clear and
efficient operational tool.

\[
\]

{\Large VII \ General case with }$N+P${\Large \ \ species competing by }$N$%
{\Large \ \ resources.-}

\[
\]

Considering the above result of section \ VI \ for two sources and two
species in competition, it seems natural to extend it to a more general
case. This will be do in this section. Consider $N$ primary resources $S_{i}$
\ ($i=1,2,...N$) and $N+P$ \ ($P\geq 0$) species $A_{j}$ \ ($j=1,2,...N+P$)
\ competing by the resources. \ We will consider this species as potential
competitors in the sense defined by expression \ (15). \ In this section,
the mean result is that at least $P$ \ species disappear. \ So, we are
considering the ecological systems given by the process 
\begin{equation}
\left\{ \sum_{i=1}^{N}S_{i}\right\} >\left\{ \sum_{j\neq k}^{N+P}\left(
A_{j}\supset \subset A_{k}\right) \right\} ,
\end{equation}
where the summation \ is understood \ in the sense of independent species in
a region, explicitly, $\sum S_{i}=S_{1}\oplus S_{2}\oplus S_{3....}$\ , \
the ecological \ process (24) is equivalent to 
\begin{equation}
(24)\Leftrightarrow \sum_{i=1}^{N}\left\{ S_{i}>\sum_{j\neq k}^{N+P}\left(
A_{j}\supset \subset A_{k}\right) \right\} ,
\end{equation}
\ because any species consumes \ any resources. Using Gause (16) for every
pair $i,k$ then we have 
\begin{equation}
(25)\Rightarrow \sum_{i=1}^{N}\left\{ S_{i}>\left( A_{1}\text{ \ or }A_{2}%
\text{ \ or }A_{3}\text{...or \ }A_{N+P}\right) \right\} \text{, \ (Gause
applied)}.
\end{equation}
So every $S_{i}$ is consumed by one species; but note that one species could
consume more than one resource. In this way, at least there are $P$ \
species extinct. The extrema \ option \ are:

\[
\]

(a) One species, called the exterminator, \ finally uses the $N$ resources.

(b) $N$ species coexist. That is, one \ species for every source (niche
differentiation).

\[
\]

{\Large VIII \ Regional and character displacement.-}

\[
\]

As said before, Gause \ competitive exclusion principle is a basic tool
which could \ be applied to more complex cases as two resources or more. In
this section \ we want to show how our notation is so coherent that it could
be applied \ to other cases. In fact we will consider displacement of
species. We will see it in a very operative way. \ 

We define displacement of a species $D$ \ from a source $S_{1}$ to $S_{2}$ as

\begin{equation}
\left\{ \left( S_{1}>D\right) \oplus \left( S_{2}>\phi \right) \right\}
\Rightarrow \left\{ \left( S_{1}>\phi \right) \oplus \left( S_{2}>D\right)
\right\} ,\text{ ( displacement).}
\end{equation}
\ \ Note that displacement could be understood \ in two ways:

(a) Regional or spatial displacements (migration). Namely, $S_{1}$ \ and \ $%
S_{2}$ \ are sources in different spatial locations.

(b) Character displacements. Namely, $S_{1}$ \ and \ $S_{2}$ \ are sources
in the same spatial place but \ species $D$ changes (displaces) its sources
necessities, for instance due to mutation.

\[
\]

Consider two specie, $A$ and $D$, competing by the same resources $S_{1}.$
Assume that $D$ displaces to the unoccupied resource $S_{2}$ before to apply
Gause. The ecological system is given by

\begin{equation}
\left\{ S_{1}>\left( A\supset \subset D\right) \right\} \oplus \left\{
S_{2}>\phi \right\} ,
\end{equation}

\begin{equation}
\Leftrightarrow \left\{ S_{1}>\left( D\supset A\oplus A\supset D\right)
\right\} \oplus \left\{ S_{2}>\phi \right\} ,
\end{equation}

\begin{equation}
\Leftrightarrow \left\{ S_{1}>D\supset A\right\} \oplus \left\{
S_{1}>A\supset D\right\} \oplus \left\{ S_{2}>\phi \right\} ,
\end{equation}

\begin{equation}
\Leftrightarrow \left\{ \left\{ \left( S_{1}>D\right) \oplus \left(
S_{2}>\phi \right) \right\} \supset A\right\} \oplus \left\{ S_{1}>A\supset
D\right\} ,
\end{equation}
where we have used $\phi \supset A\Leftrightarrow \phi $. In this stage,
assuming that species $D$ displaces to $S_{2}$ (see (27)) we have

\begin{equation}
\Rightarrow \left\{ \left\{ S_{1}>\phi \oplus S_{2}>D\right\} \supset
A\right\} \oplus \left\{ S_{1}>A\supset D\right\} ,
\end{equation}

\begin{equation}
\Leftrightarrow \left\{ S_{1}>\phi \supset A\right\} \oplus \left\{
S_{2}>D\supset A\right\} \oplus \left\{ S_{1}>A\supset D\right\} ,
\end{equation}
using newly $\phi \supset A\Leftrightarrow \phi $ \ we obtain

\begin{equation}
\Leftrightarrow \left\{ S_{2}>D\supset A\right\} \oplus \left\{
S_{1}>A\supset D\right\} .
\end{equation}
In resume, \ from (28) and (34) we have:

\begin{equation}
\left\{ S_{1}>\left( A\supset \subset D\right) \right\} \oplus \left\{
S_{2}>\phi \right\} \Rightarrow \left\{ S_{1}>A\supset D\right\} \oplus
\left\{ S_{2}>D\supset A\right\} ,
\end{equation}
\ and both species survive due to displacement. A \ practical example for
displacement comes from the same Gause classic experiments. \ In fact, when
two protozoan {\it P. caudatum} and {\it P. bursaria} were grown together
neither species suffered a decline to the point of extinction. They were in
competition with one another but although they lived together in the same
tube, they were spatially separated. {\it P. caudatum} lived suspended in
the liquid medium and {\it P. bursaria} was concentrate at the bottom of the
tube (Begon (1999). So, coexistence is related here with displacement.

An example where character displacement \ gives coexistence is provide by
mud snails \ \ {\it Hydrobia ulvae} and {\it Hydrolia ventrosa} (Saloniemi
(1993)). When they live apart, their sizes are almost identical.
Nevertheless, when put together they reach different sizes in time. In fact,
when they are similarly sized (apart) they consume similarly sized food; but
when they are put together, the more larger tends to consume larger food
particles.

\[
\]

{\Large IX Intraspecific competition : Gause does not hold.- \ }

\[
\]

As mentioned in the introduction, an important and debated question is
related to the validity of Gause's principle when high intraspecific
competition exist (individuals \ of the same species compete themselves by
resources). So, we have two species $A$ \ and $I$ exploiting a resource $S$;
but $I$ presents high intraspecific competition. In this case it seems that
both species could \ coexist (Begon (1999)). That is, assume species $A$ is
a weak consumer of $S$ and then in principle it must disappear \ face to the
strong consumer $I$; but $I$ presents a so high \ degree of intraspecific
competition that $A$ has a chance to survive.

Our formalism does not respond the question about coexistence, or not, in
this case since Gause does not hold here. In fact, we define a species $I$
with intraspecific competition as

\begin{equation}
\left( S>I\right) \Rightarrow \left( S>(I\supset I)\right) ,\text{ \ \
(intraspecific competition).}
\end{equation}
Now, consider species \ $I$ \ competing with $A$ \ by the resource $S$,
namely,

\begin{equation}
\left( S>I\supset A\right) \oplus \left( S>A\supset I\right) ,
\end{equation}
since $I$ is (high) intraspecific competitor

\begin{equation}
(37)\Rightarrow \left( S>\left( I\supset I\right) \supset A\right) \oplus
\left( S>A\supset I\right) ,
\end{equation}
and Gause does not hold since $\left( I\supset I\right) \nLeftrightarrow I$
(with exception of $\phi $). Note that we use the term high intraspecific
competition. This is so because in the above process we use \
intracompetition before applying Gause to process (37). Weak intraspecific
means that in (37) we use Gause and after the intraspecific character of
species $I$. In this case we have exclusion.

\[
\]

{\Large X Virtual process generation.- \ }

\[
\]

In this section we consider a virtual \ (speculative) possibility to
generate new processes from some known. In this sense, the new processes
constructed are not necessarily real \ process. \ The advantage of this
generation processes is to explore some future symmetries of \ ecological
systems.

We define \ the {\it dual ecological process} of a given process as this one
where the changes $\left( >\right) \rightarrow \left( \supset \right) $ and $%
\left( \supset \right) \rightarrow \left( >\right) $ \ operate. \ We define
the {\it inverse ecological process }as this one where the changes $\left(
>\right) \rightarrow \left( <\right) $\ and $\left( \supset \right)
\rightarrow \left( \subset \right) $ \ operate. For instance, the dual of $%
\left( A>B\right) $ \ denotes by $\left( A>B\right) ^{D}$ \ is $\left(
A\supset B\right) ,$ \ namely, $\left( A>B\right) ^{D}\Leftrightarrow \left(
A\supset B\right) .$ The inverse of $\left( A>B\right) $ \ is $\left(
A<B\right) $, namely \ $\left( A>B\right) ^{I}\Leftrightarrow \left(
A<B\right) $.

For instance, consider the ecological process of two species \ $A$ and $B$
in mutual depredation $(><)$ which consume also a primary resource $S$.
Moreover, since individual of every specie dies, the primary resources uses
this as a food resource (nutrients). So, consider the idealized process \
where

\begin{equation}
S>\left( A><B\right) >S,
\end{equation}
\ obviously, this process is invariant under inversion operation. Namely,

\begin{equation}
\left\{ S>\left( A><B\right) >S\right\} ^{I}\Leftrightarrow \left\{ S>\left(
A><B\right) >S\right\} .
\end{equation}

As other hypothetical example, \ consider \ the process $(S>B\supset S)$.
From the above definition for dual and inverse operations we have $%
(S>B\supset S)^{DI}\Leftrightarrow (S>B\supset S)$, namely, \ we have an
invariant ecological process under dual and inversion operations.

\[
\]

{\Large XI Time for exclusion (number of generations).-}

\[
\]

In this last section we will be concerned with a basic discussion of time
for exclusion. That is, Gause \ is a time evolutive process; but it does not
refer explicitly to how long is this time for exclusion. It seems quite
natural to think that \ when more similar species in competition are, then
more longer the extinction time becomes. \ To be more explicit, Consider two
specie \ $A$ and $B$ in competition according with Gause. Assume species $B$
is excluded in \ a number $N_{B}$ of generations. We will assume that all
similitude between both species could be quantified by one parameter $s$.
That is, $s=1$ \ both species are completely similar (same species).
Opposite, $s=0$, means that they are completely different species (genotype,
phenotype, etc.). \ The two parameter $N_{B}$ (the extinction generation
number) \ and \ the similitude parameter $s,$ are \ related by the simple
expression $N_{B}=1/(1-s),$ proposed originally in reference Flores (1998).
So, more similar the species in competition are ($s\rightarrow 1$), \ a much
longer time (number of generation) is necessary to exclusion.

\[
\]

{\Large \ \ }

{\Large XI \ Conclusions.-}

\[
\]

We have presented a \ symbology like to set theory applied to ecological
interacting process (ECOSET). Chains of depredation or competition were
explicitly \ studied. \ Particularly, Gause's exclusion principle was
considered in this notation and used as a basic operational tool. For
instance, it was applied to competition with sterile individuals (SIT), \
two species with two resources and, more general, to $N+P$ \ species with $N$
resources. The symbology is so coherent that: {\it displacement breaks
exclusion }was obtained \ with basic operations of our theory. \ Examples
from laboratory and field were explicitly considered.

\bigskip

\bigskip

\[
\]

{\Large Resume for symbols}

\[
\]

$\supset $ \ Perturbation (no depredation).

$>$ \ Depredation.

$\oplus $ \ Two independent species in a region (eventually independent
process).

$\otimes $ \ Two interdependent species.

$\Rightarrow $ Temporal evolution.

$\Leftrightarrow $ Equivalence.

$\supset \subset $ \ \ Abbreviation for competition.

$><$ \ Mutual depredation.

$or$ \ Exclusion (some times $\vee $).

$\sum $ \ Independent species (eventually process): $A\oplus B\oplus C\oplus
D\oplus ...$.

\ 

\[
\]

\end{document}